\documentclass{vgtc}                          

\ifpdf
  \pdfoutput=1\relax                   
  \usepackage{graphicx}                
  \DeclareGraphicsExtensions{.pdf,.png,.jpg,.jpeg} 
\else
  \usepackage{graphicx}                
  \DeclareGraphicsExtensions{.eps}     
\fi%

\graphicspath{{figures/}{pictures/}{images/}{./}} 

\usepackage{microtype}                 
\PassOptionsToPackage{warn}{textcomp}  
\usepackage{textcomp}                  
\usepackage{mathptmx}                  
\usepackage{times}                     
\usepackage{cite}                      
\usepackage{tabu}                      
\usepackage{booktabs}                  
\usepackage{enumitem}

\onlineid{0}

\vgtccategory{Research}

\vgtcinsertpkg

\title{Exploring Data Agency and Autonomous Agents as Embodied Data Visualizations}

\author{Sarah Schömbs\thanks{e-mail: sschombs@student.unimelb.edu.au}  %
\and Jorge Goncalves\thanks{e-mail: jorge.goncalves@unimelb.edu.au}
\and Wafa Johal\thanks{e-mail: wafa.johal@unimelb.edu.au}}
\affiliation{\scriptsize The University of Melbourne}

\teaser{
  \centering
  \includegraphics[width=\linewidth]{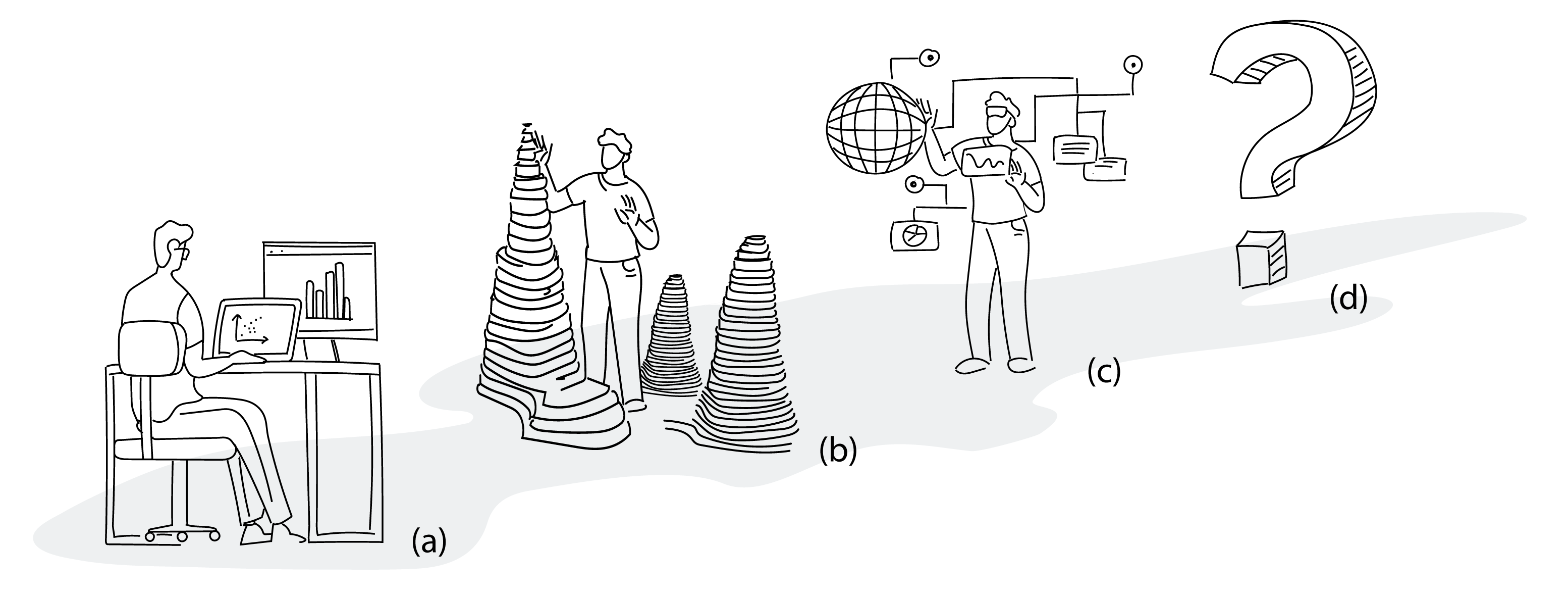}
  \caption{(a) 2D data visualization, (b) tangible physicalizations exemplified by data sculptures, (c) immersive data visualizations that can be experienced and manipulated. (d) We envision \textit{Data Agency} as a result of increasingly autonomous, interactive and adaptable visualizations.}
  \label{fig:teaser}
}

\abstract{In the light of recent advances in embodied data visualizations, we aim to shed light on agency in the context of data visualization. To do so, we introduce \textit{Data Agency} and \textit{Data-Agent Interplay} as potential terms and research focus. Furthermore, we exemplify the former in the context of human-robot interaction, and identify future challenges and research questions.} 

\CCScatlist{
  \CCScatTwelve{Human-centered computing}{Visu\-al\-iza\-tion}{Visualization theory, concepts and paradigms}{};
  \CCScatTwelve{Computer systems organization}{Embedded and cyber-physical systems}{Robotics}{External interfaces for robotics}{}
}

\begin{document}


\firstsection{Introduction}

\maketitle

Over the past decade, the data visualization community has explored various means and ways to map and visualize data. Doing so, data visualizations have gone beyond the traditional 2D graphical user interface and entered the real-world to help people explore and understand data in a more ubiquitous way, see \autoref{fig:teaser}. With the rise of applications of autonomous agents in the real-world generating data, we encourage future research to look at the \textit{Data-Agent Interplay} and \textit{Data Agency} based on two reasons: 

First, it is necessary to investigate and acknowledge potential interaction effects and perceptual changes due to the fact that data is embodied and visualized \textit{through} an agent. This \textit{Data-Agent Interplay} might become particularly important when we look at research areas that include autonomous agents and humans interacting with each other, such as human-robot interaction (HRI). Research in HRI is beginning to increasingly examine data visualizations to enhance a robot's safety \cite{safetyaura}, to visualize sensor data \cite{cleaversensar} or to convey perception results to support a robot's explainability \cite{zhaoprojection}. However, contrary to data sculptures or shape-changing bar charts, robots are perceived as agents as a consequence of their \textbf{autonomy}, \textbf{interactivity} and \textbf{adaptability} \cite{jacksonagencyhri}. Moreover, previous studies indicate that a robot's appearance, motion or behaviour affects the user's likeability, acceptance, trust and whether or not users perceive a robot as intelligent \cite{anthroroesler}. This change in perception and attitude towards robots raises several questions when it comes to data being embodied and visualized \textit{through} a robot: If data is being embodied and conveyed through a robot, how does its agency affect the interaction, user's perception and attitude towards the conveyed data? For instance, how does a robot's agency influence the user's trust in data? In line with previous research in data visualization, the latter additionally raises the question of how to map data onto a robot's behaviour, thus output parameters (e.g. light or motion). On the other hand, we also asked ourselves how visualizing data through a robot's behaviour might change the perception of and interaction with a robot (e.g. does it enhance the robot's interpretability or highlight its functionality?).

Second, the question of data agency goes beyond the HRI scope. It is arguable that agency does not have to involve explicitly designed agency, but can be ascribed, thus perceived. Agency is defined by an increase in interactivity, autonomy, and adaptability \cite{jacksonagencyhri}; so data visualizations that achieve, match and support those criteria might be perceived as an agent itself, which we define as \textit{Data Agency}. If this assumption confirms to be true, it is up to future research to inform design guidelines on how \textit{Data Agency} might look like in the future. 

After presenting the related work in the domain of data physicalization and embodiment, we propose to define \textit{Data Agency} and \textit{Data-Agent interplay} which differ from agent-enhanced data exploration and visualisation \cite{Belo2016}. Finally we draw conclusion on future research at the crossing between HRI and Data Visualization.

\section{Data Physicalization and Embodiment}
Data physicalization is a research area that investigates physical representations of data, bridging data visualization, tangible user interaction, and design. In brief, \textit{physicalizations} are shapes or forms which convey data with the aim to facilitate data exploration and analysis. Jansen et al. \cite{phyz2015} defined data physicalizations as "physical artifact[s] whose geometry or material properties encode data", earlier referred to as "beyond-desktop visualization systems" \cite{modelphyzjansen}. Contrary to immersive visualizations, physical artifacts are typically mapped physically into the real-world to enrich their perception and support their manipulation \cite{modelphyzjansen}. To illustrate, data physicalizations vary from data sculptures \cite{sculpturezhao}, physically dynamic bar charts \cite{physbarchartstaher}, to physicalizations based on swarm robots to visualize scatter plots in which each robot embodies a particular data point \cite{swarmrobots}. In the area of data physicalization, embodiment refers to the mapping of abstract data to physical representations. To do so, metaphors oftentimes serve as translators to make abstract data tangible, understandable and directly interactable \cite{sculpturezhao}. Hence, users are usually able to touch or walk through the visualized data. In the context of immersive data, embodiment is described as "the mapping of data artifacts to 3D virtual constructs that users can directly manipulate, examine, and rearrange" \cite{zhangembodiment}. Thus, embodiment in an immersive context interestingly emphasizes the user's interaction with the visualized data. Furthermore, immersive technologies, including augmented reality and virtual reality, have enabled users to experience data in situ, embedded and on-the-fly \cite{whitlock2020}.

\section{Data-Agent Interplay and Data Agency}
Data visualizations are becoming more and more experienceable through both touch and interaction. However, no prior research has investigated how data representations might affect the user's perception and interaction due to agency. We therefore introduce the term \textit{Data-Agent Interplay}, which is inspired by Satriadi et al. \cite{satriadiglobe}, and describes potential interaction effects and perceptual changes of (a) the visualized data due to the agent and (b) the agent due to the visualized data. Here, the agent does not act as an assistant or collaborator to support data exploration or analysis. Rather, the data is visualized \textit{through} the agent's behaviour, e.g. motion. We therefore define as follows:
\begin{itemize}
    \item \textit{Data-Agent Interplay describes potential interaction effects or perceptual changes based on the fact that data is visualized and represented through an agent's behaviour.}
\end{itemize}
We additionally introduce \textit{Data Agency}, which might occur if a data representation fits the description and underlying assumptions of an agent by Jackson et al.\cite{jacksonagencyhri}. We therefore define: 
\begin{itemize}
    \item \textit{Data Agency describes a phenomena that occurs when humans perceive data representations as physical or virtual agents due to their increasing interactivity, autonomy, and adaptability.}
\end{itemize}
\section{Conclusion}
The aim of this poster paper is to bring together the Information Visualization community and the HRI community \cite{szafirszafir}, to initiate a discussion in regard to \textit{Data Agency} and the \textit{Data-Agent Interplay} and to propose both as a potential focus of multidisciplinary research. Furthermore, we identified several research questions that could be addressed to explore agency in the context of data visualizations. To simplify, we exemplified agents as robots. However, it is important to highlight that \textit{Data-Agent Interplay} is not solely reduced to robots and can be expand to all sorts of embodied agents, e.g. virtual avatars.

\acknowledgments{We wish to acknowledge the contribution of the Melbourne Research Scholarship funded by the University of Melbourne. This research is partially supported by the Australian Research Council Discovery Early Career Research Award (Grant No. DE210100858). Further, we thank S. Wadinambiarachchi for her help with the visualizations.}

\bibliographystyle{abbrv-doi}

\bibliography{main}
\end{document}